\documentclass[a4paper, 12pt] {article}

\usepackage[textwidth = 430 pt, textheight = 630 pt]{geometry}
\usepackage{amsmath}
\usepackage{amssymb}
\usepackage{color}

\numberwithin{equation}{section}

\begin{document}

\pagenumbering{arabic}

\newcommand{\upu}[1]{^{\underline{#1}}}
\newcommand{\downu}[1]{_{\underline{#1}}}
\newcommand{\Epsilon}{\mathcal{E}}

\newcommand\sfrac[2]{{\textstyle\frac{#1}{#2}}}
\newcommand\ignore[1]{}
\def\one{{\,\hbox{1\kern-.8mm l}}}
\def\Sec{\S}\def\vac{|0\rangle}
\def\bra#1{\left\langle #1\right|}
\def\ket#1{\left| #1\right\rangle}
\newcommand{\braket}[2]{\langle #1 | #2 \rangle}
\newcommand{\norm}[1]{\left\| #1 \right\|}
\def\corr#1{\left\langle #1 \right\rangle}
\newcommand{\tr}{\operatorname{tr}}
\newcommand{\sech}{\operatorname{sech}}
\newcommand{\Spin}{\operatorname{Spin}}
\def\Tr{{\rm Tr\, }}
\newcommand{\Sym}{\operatorname{Sym}}
\newcommand{\SO}{\mathrm{SO}} \newcommand{\SL}{\mathrm{SL}}
\newcommand{\SU}{\mathrm{SU}} \newcommand{\U}{\mathrm{U}}
 \newcommand{\pd}{\partial}
\newcommand{\doublet}[2]{\left(\begin{array}{c}#1\\#2\end{array}\right)}
\newcommand{\twobytwo}[4]{\left(\begin{array}{cc} #1&#2\\#3&#4\end{array}\right)}

\newcommand{\cp}{\Cset \mathrm{P}}
\newcommand{\Cset}{{\,\,{{{^{_{\pmb{\mid}}}}\kern-.45em{\mathrm C}}}}}
\newcommand{\hF}{\hat F}
\newcommand{\tdr}{\tilde r}\newcommand{\cA}{\mathcal A}
\newcommand{\cB}{\mathcal B}\newcommand{\cC}{\mathcal C}
\newcommand{\cD}{\mathcal D}\newcommand{\cE}{\mathcal E}
\newcommand{\cF}{\mathcal F}\newcommand{\cG}{\mathcal G}
\newcommand{\cH}{\mathcal H}\newcommand{\cI}{\mathcal I}
\newcommand{\cJ}{\mathcal J}\newcommand{\cK}{\mathcal K}
\newcommand{\cL}{\mathcal L}\newcommand{\cM}{\mathcal M}
\newcommand{\cN}{\mathcal N}\newcommand{\cO}{\mathcal O}
\newcommand{\cP}{\mathcal P}\newcommand{\cQ}{\mathcal Q}
\newcommand{\cR}{\mathcal R}\newcommand{\cS}{\mathcal S}
\newcommand{\cT}{\mathcal T}\newcommand{\cU}{\mathcal U}
\newcommand{\cV}{\mathcal V}\newcommand{\cW}{\mathcal W}
\newcommand{\cX}{\mathcal X}\newcommand{\cY}{\mathcal Y}
\newcommand{\cZ}{\mathcal Z}
\newcommand{\hG}{{\hat \Gamma}}
\newcommand{\note}[2]{\noindent {\bf{\footnotesize [{\sc #1}}---{\footnotesize   #2]}}}
\newcommand{\ret}{\nonumber \\}
\newcommand{\nn}{\nonumber}
\newcommand{\ie}{{\it i.e.\;}}
\newcommand{\eg}{{\it e.g.\;}}
\newcommand{\be}{\begin{equation}}
\newcommand{\ee}{\end{equation}}
\newcommand{\bea}{\begin{eqnarray}}
\newcommand{\eea}{\end{eqnarray}}
\newcommand{\ov}{\over}
\newcommand{\A}{{\cal A}}
\newcommand{\C}{\mathbb C}
\newcommand{\R}{\bf R}
\newcommand{\Z}{\bf Z}
\newcommand{\bZ}{{\bar Z}}

\centerline{\LARGE \bf {\sc Charged Chiral Fermions}} \vspace{.5cm} \centerline{\LARGE \bf {\sc  from  M5-Branes }} \vspace{2truecm} \thispagestyle{empty} \centerline{
    {\large {\bf {\sc Neil Lambert}}}\footnote{ {\tt neil.lambert@kcl.ac.uk}}  
      {and} {\large {\bf{\sc Miles Owen}}}\footnote{ 
                                  {\tt miles.owen@kcl.ac.uk} }
                                                           }

\vspace{1cm}
\centerline{ {\it Department of Mathematics}}
\centerline{{\it King's College London}}
\centerline{{\it The Strand}}
\centerline{{\it London WC2R 2LS}}
\centerline{EU}

\vspace{2.0truecm}

\thispagestyle{empty}

\centerline{\sc Abstract}

We study  M5-branes wrapped on a multi-centred   Taub-NUT space. Reducing to String Theory on the $S^1$ fibration leads to  D4-branes intersecting with D6-branes. D-braneology shows that there are additional charged chiral fermions from the open strings which stretch between the D4-branes and D6-branes. From the M-theory point of view the appearance of these   charged states is  mysterious as the M5-branes are wrapped on a smooth manifold. In this paper we show how these states  arise in the M5-brane worldvolume theory and argue that  are governed by a WZWN-like model  where the topological term is five-dimensional.
  
\vfil

\setcounter{page}{0}
\newpage

\section{Introduction}

The M5-brane remains a mysterious object. For a single M5-brane the dynamical equations have   been known for some time \cite{Howe:1996yn,Perry:1996mk,Howe:1997fb, Bandos:1997ui}. At lowest order, in the decoupling limit,  these reduce to a free field theory. For $N$ M5-branes there exists an interacting CFT in six-dimensions, dubbed the $(2,0)$-theory, that captures their low energy dynamics, decoupled from gravity \cite{Witten:1995zh,Strominger:1995ac}. A reliable formulation of this theory is still lacking$(2,0)$-theory but when reduced on a circle of radius $R =  g^2/4\pi^2$ it reduces to five-dimensional maximally supersymmetric Yang-Mills (5D MSYM) with gauge group $U(N)$ and coupling $g$. Since  5D MSYM is perturbatively non-renormalizable the six-dimensional $(2,0)$ CFT provides a UV-completion with an enhanced Lorentz symmetry. It is therefore of great interest to try to understand in detail the relation of the $(2,0)$-theory to 5D MSYM. In particular one would like to know what additional states or degrees of freedom arise in the (2,0) theory that are needed to UV complete 5D MSYM. It has been suggested that  all such states are already present in 5D MSYM non-perturbatively \cite{Douglas:2010iu, Lambert:2010iw} and that 5D MSYM is in fact well-defined without new degrees of freedom.

One  case where the degrees of freedom of M5-branes seem particularly mysterious  is when we consider the $(2,0)$-theory on a multi-centred  Taub-NUT space ${\cal M}_{mTN}$. This is a completely smooth four-dimensional manifold and one expects that  the $(2,0)$-theory on ${\mathbb R}^{1,1}\times {\cal M}_{mTN}$ is locally the same as on ${\mathbb R}^{1,5}$. On the other hand reducing on the $S^1$ fibration leads to a string theory picture of $N$ D4-branes intersecting with D6-branes which are localised at the zeros of the $U(1)$ Killing vector of multi-centred  Taub-NUT space. From standard D-brane dynamics one finds that there are stretched D4-D6 strings which are localised at these zeros. In particular these are so-called `$DN=8$ strings' whose ground state consists of chiral fermions which propagate along ${\mathbb R}^{1,1}$ and lie in the bi-fundamental of $U(N)\times U(N_{I})$ where $N_I$ is the number of coincident D6-branes located at the $I^{\rm th}$ zero. These fermions have been studied in \cite{Dijkgraaf:2007sw} and  \cite{Witten:2009at}. Similar states have also  appeared in \cite{Assel:2016wcr} in the case of M5-branes wrapped on cycles in elliptic Calabi-Yau compactifications. The main question we address here is how do such charged states arise from the $(2,0)$-theory?  

This question arises even in the case of a single M5-brane, corresponding to $N=1$, where the M5-brane equations are known. However   there is still a puzzle: The chiral fermions are charged under the worldvolume gauge field  but none of the fields in the M5-brane theory have a minimal coupling so that their quanta can support a charge. This follows from the fact that for a single M5-brane all the fields have an interpretation as Goldstone modes \cite{Adawi:1998ta} and hence, by Goldstones theorem, they only have derivative interactions. We will see that the resolution of this puzzle is that the chiral fermions arise as soliton states on the M5-brane and Goldstone's theorem does not apply to solitons, {\it i.e.} Goldstone modes can have non-derivative couplings with solitons \cite{Lambert:2000np}. Aspects of this case have appeared   in \cite{Ohlsson:2012yn} and in section two we review this along with some unpublished notes \cite{LambertLiu}.

Thus the chiral  modes arise from the same sort of mechanism that appeared in  \cite{Cherkis:1997bx}. There the  chiral modes of the Heterotic string worldsheet in a ${\mathbb T}^3$ compactification were obtained from zero-modes of the 2-form gauge potential obtained in  Kaluza-Klein  reduction  of an M5-brane on $K3$. However there is a key difference here in that there is a gauge field under which the chiral modes are charged.  

In the non-abelian case of $N$ M5-branes it was argued in \cite{Witten:2009at} that the D4-D6 strings give rise to an $U(N)$ WZWN model.  The main result of this paper is to derive these states and the associated WZWN model from the $(2,0)$-theory alone, without appealing to a D-brane construction using open strings. In particular we will use a variation of 5D MSYM that was constructed in \cite{Linander:2011jy, Cordova:2013bea} as the natural non-Abelian extension of the abelian $(2,0)$-theory reduced on the circle fibration of ${\cal M}_{mTN}$. We will present these solitons in section three and obtain the WZWN model in section four. Finally in Section five we provide a  conclusion.

\section{The Abelian Case}

We start by recalling the linearized equations of motion of a single M5-brane which is just that of a six-dimensional abelian tensor multiplet \cite{Howe:1983fr} (in the notation of \cite{Linander:2011jy}):
\begin{eqnarray}
\nabla^2 \phi^{\alpha}{}_{\beta} &=&0\nonumber\\
i\Gamma^m\nabla_m\psi^\alpha &=&0\nonumber\\
H_{mnp}&=&\frac{1}{3!}\epsilon_{mnpqrs}H^{qrs}\ . 
\end{eqnarray}
Here $m,n,p=0,1,2,3,4,5$, $H_{mnp} = 3\partial_{[m}B_{np]}$ and  
$\epsilon^{012345}=1$. In addition $\alpha, \beta = 1,2,3,4$ denote indices of the fundamental \textbf{4} representation of the R-symmetry group $USp(4)$ which are raised (lowered) with the invariant tensor $M^{\alpha\beta}$ ($M_{\alpha\beta}$) and $\phi^{(\alpha\beta)}=M_{\alpha\beta}\phi^{\alpha\beta}=0$. These equations are invariant under
the supersymmetry transformations
\begin{eqnarray}
\delta \phi^{\alpha \beta} &=&
-i\bar\epsilon^{[\alpha}\psi^{\beta]}\nonumber\\ 
\delta B_{mn} &=&
-i\bar\epsilon^\alpha\Gamma_{mn}\psi_\alpha \nonumber\\ \delta\psi^\alpha &=& \nabla_m
\phi^\alpha{}_\beta\Gamma^m \epsilon^\beta + \frac{1}{2\cdot
3!}\Gamma^{mnp}H_{mnp}\epsilon^\alpha \ , 
\end{eqnarray}
provided that $\epsilon^\alpha$ is a chiral Killing spinor on
the M5-brane worldvolume: $\nabla_\mu \epsilon^\alpha=0$,
$\Gamma_{012345}\epsilon^\alpha=\epsilon^\alpha$ and subject to a reality condition. 

In our configuration the M5-brane worldvolume  is
${\bf R}^{1,1}\times {\cal M}_{mTN}$ with metric
\begin{equation}
ds^2_6 = -(dx^0)^2 + (dx^1)^2 + ds^2_{mTN}\ .
\end{equation}
Here ${\cal M}_{mTN}$ is the $n$-centred multi-centred  Taub-NUT space
\cite{Gibbons:1979xm}:
\begin{equation}
ds^2_{mTN} =
H^{-1}(dx^{5}+\theta)^2 + H d{\vec x}\cdot d{ \vec x}\ ,
\end{equation}
where
\begin{equation}
H = 1+ \sum_{I=1}^nh_I\ ,\qquad \theta
= \sum_{I=1}^n \theta_I\ ,
\end{equation}
and
\begin{equation}
h_I = \frac{R}{2}\frac{N_I}{|\vec x -\vec x_I| }\ ,\qquad
d\theta_I =  \star_3dh_I\ .
\end{equation}
For $N_I=1$ the metric is smooth
everywhere provided that one makes the identification $x^{5}\sim
x^{5}+2\pi R$. We have introduced the integer $N_I$ to allow for $N_I$ coincident D6-branes at  given pole $\vec x^I$  in the $x^7,x^8,x^9$ plane. For $N_I > 1$ this induces a conical singularity at the poles.  Asymptotically
this metric takes the form 
\begin{eqnarray}
ds^2_{mTN}
&=&
\left(1+\frac{N_{D6}R}{2r}\right)^{-1}\left(dx^{5}+\frac{1}{2}N_{D6}R\cos\theta
d\phi\right)^2 \nonumber\\ &&+\left(1+\frac{N_{D6}R}{2r}\right)\left(dr^2
+r^2d\theta^2+r^2\sin^2\theta d\phi^2\right) \ , \label{TNasym}
\end{eqnarray}
where $N_{D6}=N_1+...+N_n$ is the total number of D6-branes. In this case,
or for any other manifold $\cal M$ with self-dual curvature
 there exists a Killing spinor $\epsilon^\alpha$
that satisfies
\begin{equation}
\Gamma_{2345}\epsilon^\alpha = -\epsilon^\alpha \label{Kspin}
\end{equation}
Which is equivalent to the condition $\Gamma_{01}\epsilon^\alpha = -\epsilon^\alpha$.

Next we look for bosonic solutions to the equations of motion which
preserve all of these remaining 8 supersymmetries. Since we cannot
impose any more conditions on the Killing spinor we see that we must
have $\partial_m \phi^{\alpha}{}_{\beta}=0$. Hence without loss of generality we take
$\phi^{\alpha}{}_{\beta}=0$. Introducing light cone coordinates 
\begin{align}
x^- =
\frac{x^1-x^0}{\sqrt{2}}\qquad x^+=\frac{x^1+x^0}{\sqrt{2}}\ ,
\end{align}
we see that
\begin{equation}
\delta\psi^\alpha= \frac{1}{4}\Gamma^{-ij}H_{-ij}\epsilon^\alpha +
\frac{1}{4}\Gamma^{+ij}H_{+ij}\epsilon^\alpha + \frac{1}{2} \Gamma^{+-i} H_{+-i} +  \frac{1}{3!} \Gamma^{ijk}H_{ijk} \epsilon^\alpha=0\ ,
\end{equation}
where
$i,j=2,3,4,5$. Since $\Gamma_-\epsilon^\alpha=\Gamma^+\epsilon^\alpha=0$, and demanding the remaining 8 supersymmetries be preserved, we find
that $H_{-ij}=H_{ijk}=H_{+-i}=0$ so the solutions to the linearized equation of
motion which preserve the $(0,8)$ supersymmetries are simply
\begin{equation}
H  = \sum_{I=1}^{n}\nu_+^I dx^+\wedge \omega_I\ .
\end{equation}
Furthermore self-duality and closure of $H$ implies that  the  $\omega_I$ are self-dual harmonic 2-forms
on ${\cal M}_{mTN}$ whereas the $\nu^I_+ $ are arbitrary functions of $x^+$.

Indeed one can explicitly construct $n$ self-dual 2-forms on multi-centred  Taub-NUT  as in \cite{Ruback:1986ag}
\begin{equation}
\label{formdef} \omega_I =  \frac{1}{4\pi^2 R} d\xi_I\ ,\qquad \xi_I =
H^{-1}h_I(dx^{5} + \theta) - \theta_I\ ,
\end{equation}
where we introduce a useful normalisation to ensure that the $\omega_I$ are dimensionless and which will be justified later.
These forms are  smooth everywhere (at least in the case $N_I=1$) and satisfy
\begin{equation}
\int \omega_I\wedge \omega_J = \int \omega_I\wedge \star \omega_J =\frac{N_I}{4\pi^2} \delta_{IJ}\ .
\end{equation}  

We can also see that there are no fermion  zero-modes. In particular imposing  $\partial_-\psi^\alpha=0$ we see that the fermion equation is 
simply $\Gamma^i\nabla_i\psi^\alpha=0$ and it is a well-known result that there are no solutions to the Dirac equation which vanish at infinity. Thus the solitons  are non-degenerate and not form an enhanced multiplet of the Lorentz group.

For vanishing scalars and fermions the energy-momentum tensor is simply \cite{Maldacena:1997de}
\begin{equation}\label{norm}
T_{mn} = \frac{\pi}{2}\sqrt{-g} H_{mpq}H_n{}^{pq}\ .
\end{equation}
In which case only $T_{++}$ is non-vanishing and we define
\begin{eqnarray}
{\cal P}_+ 
&=& \int d^5x \, T_{++} \nonumber \\ &=& \frac{1}{4\pi} \sum N_I \int dx^+ \nu^I_+(x^+) \nu^I_+(x^+) \ .
\end{eqnarray}

In particular the abelian $(2,0)$-theory contains the conserved current (we choose the coefficient for future convenience)
\begin{equation}\label{Jis}
J_m (\Lambda)  = 2\pi \sqrt{-g}H_{mnp}\partial^n \Lambda^p\ ,
\end{equation}
for any choice of 1-form $\Lambda$ inherited from the gauge symmetry $B\to B+d\Lambda$. On-shell the associated charge is a total derivative:

\begin{eqnarray}
{\cal Q}(\Lambda) &=& \int_{{\mathbb R}\times{\cal M}_{mTN}} J_+(\Lambda) d^4x dx^+\nonumber\\
&=&2\pi\oint_{{\mathbb R}\times S^1\times{S^2_\infty}} H_{+r\mu} \Lambda^\mu \, r^2d\Omega_2 dx^+\ ,
\end{eqnarray}
where $S^1\times S^2_\infty$ is the asymptotic form of ${\cal M}_{mTN}$ and $r$ the radial direction. Taking only $\Lambda_5$ non-vanishing we find
\begin{eqnarray}
{\cal Q}(\Lambda_{5}(\infty) ) &=& \frac{1}{2\pi R}\text{tr}\sum_I \oint_{{\mathbb R}\times S^1\times{S^2_\infty}}  d\Omega_2  dx^+ \, \left[  H\partial_r \left(\frac{h_I}{H}\right) + \varepsilon^{rjk}  \theta_j \partial_k \left(\frac{h_I}{H}\right)  \right] \nu_+^I \Lambda_5(\infty)\nonumber \\
&=& -  2\pi R\sum_I N_{I}\int dx^+  \nu_+^I(x^+)\Lambda_{5}(\infty)\ ,
\label{chargeA}
\end{eqnarray}
where the second term in the first line arises as $\Lambda^i = g^{i5}\Lambda_5\ne 0$. 
Upon reduction on the $S^1$ parameterized by $x^5$ the D4-brane $U(1)$ gauge field is $A_\mu = 4\pi^2 R B_{\mu 5}$ \cite{Hull:2014cxa} and the $U(1)$ gauge symmetry is $A_\mu\to A_\mu + 4\pi^2 R\partial_\mu\Lambda_{5}$. Thus ${\cal Q}(\Lambda_{5}(\infty))$ is the corresponding  electric charge that we are looking for and each $\nu^I_+$ carries $N_I$ units of its charge.

\section{The Non-Abelian Case}

In general there is no satisfactory formulation of the M5-brane in the non-Abelian case. Nevertheless the M5-brane on a circle of radius $R$ gives, at least at low energy, 5D MSYM. Therefore one can reduce the abelian theory on the $S^1$ fibration in ${\cal M}_{mTN}$ and then find the appropriate non-abelian generalisation. This was done in \cite{Linander:2011jy, Cordova:2013bea}. Let us first give their result. Reducing on $x^{5}$ leads to the five-dimensional metric
\begin{equation}
ds^2_5 = -(dx^0)^2 + (dx^1)^2 + Hd{\vec x}\cdot d{\vec x}\ .
\end{equation}
For our purposes we need that the gauge field action is\footnote{We use a convention where   $\frac{1}{8\pi^2 }{\rm tr}\int F\wedge F\in{\mathbb Z}$.}
\begin{eqnarray}
S_F &=&   \frac{1}{8\pi^2 R} \int d^5x \sqrt{H}   {\rm tr}(F\wedge \star F)  +  \theta\wedge {\rm tr}(F\wedge F)  \ , \label{S1}
\label{action}
\end{eqnarray}
where $\mu,\nu=0,1,2,3,4$. 
For computing the energy-momentum tensor we will also need the scalar action which is 
\begin{eqnarray}
S_\phi &=& -\frac{1}{8\pi^2 R} {\rm tr}\int d^5x  \sqrt{-g} \left( \sqrt{H} D_\mu \phi_{\alpha \beta} D^\mu \phi^{\alpha \beta}  +\frac{1}{4} \frac{1}{H^{5/2}} \partial_i H \partial_i H \phi_{\alpha \beta} \phi^{\alpha \beta}\right.\nonumber\\
&&\left. \qquad\qquad\qquad \qquad\qquad   - \sqrt{H}[\phi^{\alpha \beta}, \phi_\beta{}^\delta ] [\phi_{\delta \gamma}, \phi^\gamma{}_\alpha]  \right) \ .
\end{eqnarray}
Note that we could introduce an alternative form for the gauge part of the action:
\begin{eqnarray}
S'_F
&= &  \frac{1}{8\pi^2 R} \int d^5x \sqrt{H}   {\rm tr}(F\wedge \star F)  +  {\cal F}\wedge  CS\ ,\label{S2} 
\label{action}
\end{eqnarray}
where
\begin{equation}
 CS  =   {\rm tr}\left (A_{\mu}\partial_\nu A_{\lambda} + \frac23 A_{\mu}A_\nu A_{\lambda}\right)dx^\mu\wedge dx^\nu\wedge dx^\lambda\ .
\end{equation}
These two actions differ by whether the topological term is taken to be $\theta\wedge {\rm tr}({F\wedge F})$ or ${\cal F}\wedge CS$. In turn these  choices differ by boundary terms arising from the poles of $H$ and infinity and hence have the same equations of motion. The  first choice preserves all gauge symmetries of the action but depends upon the choice of $\theta$ and hence is not diffeomorphism invariant. Whereas the second form is diffeomorphism invariant but at the expense of introducing potential violations of worldvolume gauge symmetries. We will mainly be interested in the first case, however in section four we will explore some of the physical differences that arise from the second and which rule it out as the correct one.  Indeed part of the motivation of this paper is to explore such subtleties.
  
\subsection{D4-D6 Strings as Solitons}

We work from results in \cite{Linander:2011jy} which give the 5D theory resulting after reduction over $x^5$. The prescription for the decomposition from 6D to 5D is given in the paper and we thus denote the decomposed 5D gamma matrices by $\gamma$, and the 5D Killing spinor by $\varepsilon$. One then finds that equation \eqref{Kspin} reduces, after the decomposition, to the condition

\begin{equation}
i \gamma_{234}\varepsilon^\alpha  = \varepsilon^\alpha\ ,
\label{5dcond1}
\end{equation}
equivalently
\begin{equation}
\gamma^{01} \varepsilon^\alpha = \varepsilon^\alpha\ .
\label{5dcond2}
\end{equation}
The fermionic supersymmetry variation from \cite{Linander:2011jy} is given by
\begin{eqnarray}
\delta \psi^\alpha &=&\frac{1}{2} F_{\mu \nu} \gamma^{\mu \nu} \varepsilon^\alpha + 2i \sqrt{H} M_{\beta \gamma} D_\mu \left( \frac{1}{\sqrt{H}} \phi^{\alpha \beta} \right)\gamma^{\mu} \varepsilon^{\gamma} \nonumber \\ &-& \frac{1}{\sqrt{H}}M_{\beta \gamma} \phi^{\alpha \beta} \mathcal{F}_{\mu \nu} \gamma^{\mu \nu} \varepsilon^{\gamma} + 2 M_{\beta \gamma} M_{\delta \lambda} [\phi^{\alpha \beta}, \phi^{\gamma \delta}] \varepsilon^\lambda\ ,
\label{fervar}
\end{eqnarray}
with $\mathcal{F} = d\theta$ and we recall that the 6 dimensional two form, $B_{\mu \nu}$, is reduced to a $U(1)$ gauge field as $A_\mu = B_{\mu 5}$ with corresponding field strength 
\begin{equation}
F_{\mu \nu} = \nabla_\mu A_\nu - \nabla_\nu A_\mu + [A_\mu, A_\nu]\ ,
\end{equation}
and thus a gauge covariant derivative defined by
\begin{equation}
D_\mu \chi = \nabla_\mu \chi + [A_\mu, \chi]\ ,
\end{equation}
where is $\chi$ some field transforming in the adjoint of the gauge group. 

We seek bosonic, BPS states of the configuration to find those maximally supersymmetric states. This is equivalent to setting equation \eqref{fervar} to zero. Using the Killing spinor conditions above and after changing to the light cone coordinates introduced in the abelian case, we find that the BPS conditions for this system are 
\begin{equation}\label{BPS}
F_{ij} = F_{+-}=F_{i-}=0\ ,
\end{equation}
where from now on $i,j=2,3,4$ and also
\begin{equation}
D_i \left( \sqrt{H}\phi^\alpha{}_\beta \right) =D_-\phi^\alpha{}_\beta =0 , \qquad [\phi^\alpha{}_\beta,\phi^\beta{}_\gamma]=0\ .
\label{scalarbps}
\end{equation}
In addition one can compute the equation of motion from the action \eqref{action} and obtain
 \begin{equation}
\sqrt{-g} D_\sigma \left( \sqrt{H} F^{\sigma \lambda} \right)+ \frac{1}{4} {\cal F}_{\mu \nu} F_{\rho \sigma} \epsilon^{\mu \nu \rho \sigma \lambda} = 0\ .
\end{equation}
Upon enforcing the BPS conditions above this equation of motion reduces to
\begin{equation}
\partial_i F_{+i} + [A_i, F_{+i}]+2\partial _iH F_{+i}=0 \ .
\label{Feom}
\end{equation}
First, looking at \eqref{BPS}, we choose to set $A_i=A_-=0$, then we have that $A_+ = A_+(x^+, x^i)$ solves these conditions.

Now turning to \eqref{scalarbps}, notice that a solution is given by the ansatz $\phi^\alpha{}_\beta = \frac{1}{\sqrt{H}} \phi{}^\alpha_0{}_\beta(x^+)$ with the understanding that $[\phi^\alpha_0{}_\beta , \phi^\beta_0{}_\gamma] = 0$.

To solve the equation of motion \eqref{Feom}  we start by noting that the general solution to the BPS conditions $F_{ij}= F_{i-}=0$ is given by
\begin{equation}\label{Agg'}
A_i = g\partial_i g^{-1}\qquad A_- = g\partial_-g^{-1}\ ,
\end{equation}
for a arbitrary element $g$ of the unbroken gauge group. Similarly the solution to BPS condition $F_{+-}=0$ implies that
\begin{equation}
A_+ = g'\partial_+ g'^{-1}\qquad A_- = g'\partial_-g'^{-1}\ ,
\end{equation}
for some other element $g'$ of the unbroken gauge group. Consistency of these two expressions for $A_-$ implies that $g'^{-1}g\partial_-(g^{-1}g')=0$ and hence
\begin{equation}\label{genA}
g'=   gk\qquad {\rm with }\qquad \partial_-k=0 \ .
\end{equation}
 Thus we see that the generic solution to the BPS equation is simply a gauge transformation  by $g$ of the configuration  $A_+ = k\partial_+ k^{-1}$, $A_-=A_i=0$, corresponding to  $F_{i+} = \partial_i A_+$.

To continue then we fix the gauge $A_-=A_i=0$ and pick an ansatz for $A_+$ of the form $A_+ = K(\vec x)\nu_+(x^+)$ for some $K(\vec{x})$; this means that equation of motion becomes  
\begin{equation}
\partial_i\partial_i K + \frac{2}{H}\partial_iK\partial_i H = 0\ .
\end{equation}
Solutions to this equation are of the form
\begin{equation}
K = \frac{h}{H}\ ,
\end{equation}
where $h$ is any harmonic function: $\partial_i\partial_i h=0$. However, we wish to look for solutions with finite energy. To achieve this, any pole in $h$ must be cancelled by a pole in $H$  (see the expressions below for the energy-momentum tensor) and therefore we find the solutions
\begin{equation}
K_I  = \frac{h_I}{H}  = \frac{h_I}{1+\sum_J h_J}\ .
\end{equation}

One might worry that there is another finite energy solution $K_0$ corresponding to $h=1$. However one sees that
\begin{equation}
\sum_I K_I = \frac{H-1}{H} = 1-K_0\ .
\end{equation}
Rearranging this we see that the solution
\begin{equation}
A_+ = K_0\nu^0_+ +\sum_I K_I\nu^I_+ = \nu^0_+ +\nu^1_+ +...+\nu^n_+ \ ,
\end{equation}
is pure gauge. 
Therefore we conclude that the 
most general finite-energy soliton solution is 
\begin{equation}
A_+ = \sum_{I=1}^n K_I(\vec x)\nu^I_+(x^+)\ ,
\end{equation}
where $\nu^I_+$ is an arbitrary $x^+$-dependent element of the unbroken gauge algebra. Of course one can indeed check that these functions $K_I$ also appear in  the self-dual 2-forms constructed above as  $K_I = \xi_{I5}$. In particular our solutions are  
\begin{eqnarray}
F &=&  \sum_I \nu^I_+(x^+)\partial_iK_I dx^+\wedge dx^i \nonumber\\
&=& 4\pi^2 R\sum_I \nu^I_+(x^+) \omega^I_{i5} dx^+\wedge dx^i \ ,
\end{eqnarray}
which corresponds to a simple embedding of the abelian solution into the non-abelian theory by promoting $\nu^I_+$ to a element of the unbroken M5-brane gauge algebra and identifying 
\begin{equation}\label{FH}
F_{\mu\nu} =4\pi^2 RH_{\mu\nu 5}\ ,
\end{equation}
in agreement with \cite{Hull:2014cxa}, explaining our normalization in (\ref{norm}).

We can also see that there are no fermionic zero-modes. The fermionic equation is \cite{Linander:2011jy}
\begin{equation}
i\sqrt{H}\gamma^\mu D_\mu\psi^\alpha - \frac{1}{8}{\cal F}_{\mu\nu}\gamma^{\mu\nu}\psi^\alpha =0\ .
\end{equation}
Imposing  $\partial_-\psi^\alpha=0$ and expanding around our solitons we find this splits into two chiral equations
\begin{eqnarray}
-\sqrt{2}\gamma_0 H^{\frac12}D_+\psi^\alpha_+ +\vec\gamma\cdot\vec \nabla\psi^\alpha_- +\frac14 H^{-\frac12}\vec\gamma\cdot\vec\nabla H \psi^\alpha_- &=& 0 \nonumber\\
\vec\gamma\cdot\vec \nabla \psi^\alpha_+ -\frac14 H^{-\frac12} \vec\gamma\cdot\vec\nabla H \psi^\alpha_+ &=& 0 \ .
\end{eqnarray}
Note that the only appearance of the non-abelian gauge field is through the $D_+$ term in the first equation. 
The second equation is simply the Dirac equation for $\hat\psi^\alpha_+ = e^{-\frac12 H^{1/2}}\psi^\alpha_+$, {\it i.e.} $\vec\gamma\cdot\vec \nabla \hat\psi^\alpha_+=0$. As with the abelian case there are no solutions which vanish at infinity and hence $\psi^\alpha_+=0$. In this case the first equation becomes the Dirac equation   $\vec\gamma\cdot\vec \nabla \hat\psi^\alpha_-=0$ where $\hat\psi^\alpha_- = e^{\frac12 H^{1/2}}\psi^\alpha_-$ and we again conclude that $\psi^\alpha_-=0$. 
Thus the solitons do not form enhanced representations of the Lorentz group.

It is useful to note that, in terms of the group element $k$ defined by $A_+=k\partial_+k^{-1}$, we have
\begin{equation}
k^{-1} = Pexp\left(\sum_IK_I(\vec x )\int^{x^+}_0\nu^I_+(y^+)dy^+\right)\ .
\end{equation}
Furthermore we observe that $K_I(\vec x_J) = \delta_{IJ}$
and hence
\begin{equation}\label{kisv}
A_+(\vec x_I)=k(\vec x_I)\partial_+k^{-1}(\vec x_I) = \nu^I_+(x^+)\ .
\end{equation}
Thus although the gauge fields are spread-out over the whole of the multi-centred  Taub-NUT space there is a sense in which the chiral mode $\nu^I_+$ is associated to the $I$-th pole in $H$. Furthermore far from the poles the field strength falls-off as $1/|\vec x|^2$ as expected for a massless charged particle in $4+1$ dimensions. However it is amusing to observe that near a pole $\vec x^I$ the gauge field
\begin{equation}
A_+\sim  \frac{RN_I/2}{RN_I/2+|\vec x-\vec x_I|}  \nu^I_+(x^+)\ ,
\end{equation}
is finite \cite{LambertLiu}. In particular for $|\vec x-\vec x_I|>> R$ the solution can be written terms of an infinite expansion  of  perturbative $g^2=4\pi^2 R$ corrections to the  familiar $g^2/4\pi^2|\vec x-\vec x_I|$ Coloumb potential. 

The energy-momentum tensor, $T_{\mu \nu} = \frac{-2}{\sqrt{-g}} \frac{\delta \mathcal{L}}{\delta g^{\mu \nu}}$, is readily found to be 
\begin{eqnarray}
T_{\mu \nu} &=&   \frac{1}{8\pi^2 R}{\rm tr}\Big[2\sqrt{H} D_\mu \phi_{\alpha \beta} D_\nu \phi^{\alpha \beta} + \frac{1}{2H^{3/2}} \partial_\mu H \partial_\nu H \phi_{\alpha \beta} \phi^{\alpha \beta} + 2 \sqrt{H} F_{\mu \rho} F_\nu{}^\rho \nonumber \\ &\phantom{=}&
\qquad\qquad  - g_{\mu \nu} \Big( \sqrt{H} D_\rho \phi_{\alpha \beta} D^\rho \phi^{\alpha \beta} + \frac{1}{4} \frac{1}{H^{5/2}} \partial_i H \partial_i H \phi_{\alpha \beta} \phi^{\alpha \beta} +\frac{\sqrt{H}}{2}F_{\rho \sigma} F^{\rho \sigma} \nonumber \\ &\phantom{=}& \quad \phantom{-g_{\mu \nu} \Big(}- \sqrt{H}[\phi^{\alpha \beta}, \phi_{\beta \lambda}][\phi^{\lambda \rho}, \phi_{\rho \alpha}]  \Big)\Big] \ .
\end{eqnarray}
So that on our solution
\begin{eqnarray}
 T_{++} &=& \frac{1}{4\pi^2 R}\frac{1}{\sqrt{H}} \left(  D_+ {\phi_0}_{\alpha \beta} D_+ \phi_0^{\alpha \beta} + \sum_{IJ}\partial_i K_I \partial_i K_J \nu^I(x^+) \nu^J(x^+)   \right) \nonumber \\ T_{+-} &=& -\frac{1}{32 \pi^2 R} \frac{1}{H^{7/2}} \partial_i H \partial_i H {\phi_0}_{\alpha \beta} \phi_0^{\alpha \beta} \nonumber \\ T_{i+} &=& -\frac{1}{8 \pi^2 R} \frac{1}{H^{3/2}} \partial_i H {\phi_0}_{\alpha \beta} D_+ \phi^{\alpha \beta}_0\ .
\end{eqnarray}
Finiteness of the energy-momentum tensor implies that $D_+\phi^\alpha{}_\beta=0$. This is satisfied easily by demanding $\phi^\alpha_0{}_\beta$ be a constant, in particular we pick $\phi^\alpha_0{}_\beta = 0$ so that the unbroken gauge algebra is ${\mathfrak u}(N)$. With this extra step the energy momentum tensor again reduces to a very simple form where only $T_{++}$ is non-zero and is given by
\begin{equation}
T_{++} = \frac{1}{4\pi^2 R}\frac{1}{\sqrt{H}}{\rm tr} \sum_{IJ}\partial_i K_I \partial_i K_J \nu^I(x^+) \nu^J(x^+)\ .
\end{equation}
We then proceed to  explicitly compute the integral over the internal $\mathbb{R}^3$ to find
\begin{eqnarray}\label{P+}
{\cal P}_+ &=&\int \, d^3xdx^+ \sqrt{-g} \,  T_{++} \nonumber \\
  &=&   \frac{1}{4\pi} \sum_I N_I {\rm tr}\int dx^+  \nu^I(x^+) \nu^I(x^+)\ .
\end{eqnarray}
This  agrees with the abelian case above. 
Furthermore we see that  (\ref{P+}) corresponds 
precisely to $n$ copies, where $n$ is the number of centres of  ${\cal M}_{mTN}$,  of   a WZWN model each at level $N_I$. However given that the value of $N_I$ can be different for each $I$ we can't simply use a standard WZWN model on a three-manifold with $n$ boundaries. We will return to this issue in the next section.

Next we look at the gauge charges. For the first form of the action (\ref{S1}) we find
\begin{eqnarray}
J^\sigma(\Lambda) &=& 
\frac{1}{8 \pi^2 R} \text{tr} \left[ -2 \sqrt{-g} \sqrt{H} F^{\sigma \lambda} D_\lambda \Lambda + \varepsilon^{\mu \nu \rho \sigma \lambda} \theta_\mu F_{\nu \rho}  D_\lambda \Lambda \right]
\nonumber\\
&=&\frac{1}{4\pi^2 R} \partial_\lambda \text{tr} \left( \sqrt{-g} \sqrt{H} F^{\lambda \sigma } \Lambda   + \frac{1}{2} \varepsilon^{\mu \nu \rho \sigma \lambda}     \theta_\mu F_{\nu \rho} \Lambda  \right)    \nonumber \\ 
&\phantom{=}& -{\frac{1}{4\pi^2 R} }   \text{tr}\left( \sqrt{-g} D_\lambda\left( \sqrt{H} F^{\lambda \sigma} \right) + \frac{1}{4} \varepsilon^{\mu \nu \rho \lambda \sigma} \mathcal{F}_{\mu \nu} F_{\rho \lambda} \right)  \ ,
\end{eqnarray}
where the last line vanishes on-shell. 
The associated charges are \begin{eqnarray}
\mathcal{Q} (\Lambda(\infty))&=& \frac{1}{4 \pi^2 R} \text{tr}\sum_I \oint d\Omega_2  dx^+ \, \left[ H \partial_r K_I + \varepsilon^{rjk} \theta_j \partial_k K_I   \right] \nu_+^I \Lambda(\infty)\nonumber\\
&=& -\frac{1}{2 \pi } \text{tr} \sum_IN_I \int  dx^+ \,  \nu_+^I \Lambda(\infty)\ ,
\end{eqnarray}
where $\Lambda(\infty)$ is any element of the  unbroken gauge algebra.
These charges only receive contributions from infinity and as such do not depend on the choice of $\theta$. We see that they are the natural non-abelian extension of (\ref{chargeA}) with the identification $\Lambda = 4\pi^2 R \Lambda_5$.

\section{Gauge Symmetries and a WZWN-like Action}

As we mentioned above there are two choices for the five-dimensional action. The results in the previous section correspond to the first choice (\ref{S1}). In this section we wish to explore some physical consequences of the other choice of the action (\ref{S2}). We will then use this analysis to motivate a WZWN-like model as the effective action for the chiral soliton modes found above. 

\subsection{Physical `Gauge' Transformations}

The main difference between the two forms for the action can be seen from their gauge symmetry. While the first form is gauge invariant the second is not. In particular  the second form of the action (\ref{S2})  transforms as (assuming boundary conditions that allow us to ignore boundary terms in $x^+$) 
\begin{eqnarray}
\delta_\Lambda S &=&  -\frac{1}{4\pi} N_{D6}\int d^2x {\rm tr}\left((\partial_+A_-(\infty)-\partial_-A_+(\infty))\Lambda(\infty)\right)\nonumber\\
&& \qquad\qquad +\frac{1}{4\pi}\sum_I N_I\int  d^2x{\rm tr}\left((\partial_+A_-(\vec x_I)-\partial_-A_+(\vec x_I))\Lambda(\vec x_I)\right)\ .
\end{eqnarray}
We can make the first line vanish by imposing a suitable boundary condition at infinity. However for the other terms it seems more natural to restrict the gauge symmetry so that 
\begin{equation}
\Lambda(\vec x_I) =0\ .
\end{equation}
As we will see this has the effect of introducing additional degrees of freedom that live at the poles $\vec x_I$. These arise because there are now  transformations  of the soliton solution generated by $\Lambda(\vec x_I)$ which  lead to physically distinct states.

To continue we   evaluate the action  (\ref{S2}) on the full space of BPS solutions, including dependence of $g$ on $x^+,x^-, \vec x$. The first term of the action is still vanishing. However 
substituting the general ansatz (\ref{Agg'})-(\ref{genA}) into the second form of the action (\ref{S2}) we find
\begin{eqnarray}
S_{BPS} &=&  \frac{1}{8\pi^2 R}{\rm tr}\int {\cal F}\wedge (A\wedge dA+\frac23 A\wedge A\wedge A)\nonumber\\
&=& \frac{1}{8\pi^2 R}  \int \partial_i H (CS)_{+-i} dx^+ dx^- d^3x\ .
\end{eqnarray}
Evaluating the action on our BPS sector gives
\begin{eqnarray}\label{SBPS}
S_{BPS} 
&=&  \frac{1}{8\pi^2 R} {\rm tr}\int \partial_i H ( \partial_i(A_+A_-)+A_i\partial_+A_--A_-\partial_+A_i ) dx^+ dx^- d^3x\ ,
\end{eqnarray}
where we have used the fact that $F_{i-}=0$ and assumed boundary conditions along $x^-$ that allow us drop boundary terms in $x^-$. 

There are two ways to proceed. The first is analogous to the classic construction of \cite{Elitzur:1989nr}. In that treatment one integrates over the $A_+$ gauge field which imposes the  constraint  $F_{-i}=0$. Here  we do not integrate  over $A_+$. Rather we have imposed the BPS conditions, which  includes the constraint $F_{-i}=0$,  and evaluated the action. To this end we 
 integrate the first term  in (\ref{SBPS}) by parts and, observing that
 \begin{equation}
\partial_i\partial_i H = -2\pi \sum N_IR\delta^3(\vec x-\vec x_I)\ ,
\end{equation}
we find a contribution
\begin{equation}\label{bterm}
S_{BPS} =  \frac{1}{4\pi  }\sum_IN_I {\rm tr}\int  dx^+ dx^-   A_+(\vec x_I)A_-(\vec x_I) +...\ .
\end{equation}
To continue in analogy with \cite{Elitzur:1989nr} we assume a boundary condition such that $A_+(x_I)=0$ for each $I$. With this condition the full action reduces to:
\begin{eqnarray}
S_{BPS} &=& -\sum_I \frac{N_I}{4\pi} {\rm tr} \int dx^+ dx^- \, g(\vec{x}_I) \partial_+ g^{-1}(\vec{x}_I)  g(\vec{x}_I) \partial_- g^{-1}(\vec{x}_I) \nonumber \\ &\phantom{=}& + \frac{1}{8 \pi^2 R} {\rm tr} \int d^5x \, \partial_i H [g^{-1}\partial_-g, g^{-1}\partial_+g] \, g^{-1} \partial_i g\ .
\label{bpsaction}
\end{eqnarray}
This is essentially a WZWN model  with $n$ two-dimensional `boundaries' located at the poles of $H$ each with level $N_I$ (although we recall that only $N_I=1$ corresponds to a completely smooth multi-centred  Taub-NUT space). The difference with a traditional WZWN model is that in our case the topological term is five-dimensional and the two-dimensional `boundary' contributions arise from the poles of $H$. Nevertheless it plays the same role as the familiar three-dimensional term. In particular the associated equation of motion is restricted to the poles and is given by
\begin{equation}
\partial_+( g(\vec{x}_I)\partial_-g^{-1}(\vec{x}_I))=0\ ,
\end{equation}
for each $I$.
We thus obtain a theory of $n$ independent two-dimensional group-valued fields $g(\vec x_I)$.
The solution to this is simply
\begin{equation}
g(\vec{x}_I) =  \ell_I(x^-)r_I(x^+)\ .
\end{equation}
for arbitrary group elements $\ell_I(x^-)$ and $r_I(x^+)$.
 However   we must ensure that the boundary condition $A_+(x_I)=0$ is satisfied. One finds that this implies 
 \begin{equation}
 r_I = k^{-1}(\vec{x}_I)\ ,
 \end{equation}
and hence
\begin{equation}
g(\vec{x}_I) =  \ell_I(x^-)k^{-1}(\vec x_I)\ .
\end{equation}
Thus we are left with a single independent group element $\ell_I(x^-)$ in addition to the original solution $k^{-1}(\vec x_I)$ 

The second approach is  to include the `boundary'  term (\ref{bterm}) into the action which we again evaluate on the BPS solutions, {\it i.e.} we do not impose any conditions on $A_+$ at the poles. In this case find
 \begin{eqnarray}
S_{BPS} &=&   \sum_I \frac{N_I}{4\pi} {\rm tr} \int dx^+ dx^- \, k(\vec{x}_I) \partial_+ k^{-1}(\vec{x}_I)  \partial_- g^{-1}(\vec{x}_I) g(\vec{x}_I) \nonumber \\ &\phantom{=}&+
\frac{1}{8\pi^2 R} {\rm tr} \int d^5x \, \partial_i H [g^{-1}\partial_-g, g^{-1}\partial_+g] \, g^{-1} \partial_i g\ .
\label{bpsaction2}
\end{eqnarray}
Here the standard quadratic kinetic term for $g$ has been removed and replaced by a linear term coupled to the background field $k$.
The associated equation of motion still only receives contributions from the poles but has a less familiar form:
\begin{eqnarray}\label{EoM2}
0&=& \partial_-g k(\vec x_I)\partial_+k^{-1}(\vec x_I)g^{-1} +gk(\vec x_I)\partial_+ k^{-1}(\vec x_I)\partial_- g^{-1} \nonumber\\
&& + g\partial_+g^{-1}g\partial_-g^{-1} - g\partial_-g^{-1} g\partial_+g^{-1}\ ,
\end{eqnarray}
for each $I$. To solve this we can write 
\begin{equation}
g(\vec x_I) = \ell_I(x^+,x^-) k^{-1}(\vec x_I)\ ,
\end{equation}
for some $\ell_I$ that is now allowed to depend on both $x^-$ and $x^+$. Substituting this into (\ref{EoM2}) we simply find, for each $I$, 
\begin{equation}
\left[ \ell_I \partial_+ \ell^{-1}_I, \ell_I \partial_- \ell^{-1}_I\right] = 0\ .
\end{equation}
There are essentially two  ways to satisfy this equation. Firstly, if $\ell_I \partial_- \ell^{-1}_I = 0$ then we have  $\ell_I = \ell_I(x^+)$. This means that $g = \ell_I k^{-1}(\vec x_I)$ is a function only of $x^+$ and hence $\ell_I$ can be absorbed into a redefinition of $\nu_I(x^+)$.  
The second solution is to demand $\ell_I \partial_+ \ell^{-1}_I = 0$ so we have $\ell_I = \ell_I(x^-)$. In this case we recover the same solutions that we saw above by imposing the vanishing of $A_+(\vec{x}_I)$.

In summary we find that with the second choice of action (\ref{S2})  there are some gauge modes which are physical. In particular we find that the solution space includes the modes $\ell_I(x^-)$  that arises from the broken  gauge  modes. Hence we can think of it as a physical Goldstone mode and the WZWN-like model as its low energy effective action. However we do not expect such modes to arise from the D-brane analysis and hence we conclude that (\ref{S2}) is the wrong choice of action.

\subsection{An Action for the Soliton Modes}

We now return to the original action (\ref{S1}). Here we can simply adapt the argument above. 
We have seen that the D4-D6 strings can be realised in the non-Abelian theory as solitons. We have evaluated their energy and momentum and shown that they agree with that of a chiral half of a WZWN model.  To capture the effective dynamics of these solitons we therefore propose that the action (\ref{bpsaction}) can be used with a slightly modified interpretation. In particular we recall that the solution to the equations of motion can be written as 
\begin{equation}
g(\vec{x}_I) = \ell_I(x^-)r_I(x^+)\ ,
\end{equation}
for arbitrary left and right moving modes $\ell_I$ and $r_I$. To make contact with our solitons we first set $\ell_I$ to the identity and identify
\begin{equation}
\nu^I_+(x^+) = r_I(x^+) \partial_+ r^{-1}_I(x^+)  \ .
\end{equation}
{\it i.e.} $r_I(x^+) = k(\vec{x}_I)$ in (\ref{kisv}). We also see that taking a non-trivial $\ell_I(x^-)$ can be viewed as performing the gauge transformation: $A_+(\vec{x}_I) = \ell k\partial_+(k^{-1}\ell^{-1})$ and $A_-(\vec{x}_I)= \ell\partial_-\ell^{-1}$.  Therefore we consider the other chiral half to be pure gauge and we simply discard it. This is consistent with the discussion above where such gauge modes were physical and therefore not discarded.

\section{Conclusions}
 
 In this paper we have studied how the charged D4-D6 strings which arise from a D4-brane intersecting with a D6-string are realised in the M5-brane worldvolume theory. In particular we showed that there are smooth soliton solutions of the five-dimensional Yang-Mills gauge theory arising from the M5-brane reduced on the circle fibration of multi-centred  Taub-NUT space that have the right charges to be identified with the D4-D6 strings. We also considered the physical consequences of the two choices of action and how the second choice leads to additional physical soliton zero-modes which do not match the string theory analysis. Lastly we obtained a WZWN-like model for the solitons but where the topological term is five-dimensional. We thus conclude that 5D MSYM contains the charged states predicted from the D-brane construction, albeit as solitons. 

Let us briefly mention some bulk eleven-dimensional aspects of our solutions.
The states we have identified arise as stretched D4-D6-strings. In the string theory picture these states are localized to the intersection. In M-theory they lift to M2-branes that wrap the M-theory circle. Since the M-theory circle shrinks to zero at the poles of $H$ the M2-brane worldvolume theory develops a potential $V\propto H^{-1/2}$ and so the  energy is minimized by sticking to the poles ${\vec x}_I$, in agreement with the microscopic string theory picture. 

Our solutions are given in terms of harmonic forms which can also be associated to the existence of non-trivial two-cycles in multi-centred Taub-NUT. These two-cycles are caused by the shrinking of the circle fibration at the poles of $H$ and so can be thought of as connecting two distinct poles.  M2-branes wrapping these cycles are in bi-fundamental representations of $U(1)\times U(1)$ subgroups of a $U(1)^{N_{D6}-1}$ gauge group whose bulk gauge field arises from a Kaluza-Klein reduction  of the M-theory three-form $C\sim \sum C_I\wedge \omega_I$ (here we are  one thinking of multi-centred Taub-NUT as compact or replacing it by a similar compact space).  When all the D6-branes coalesce this group is enhanced  and the wrapped M2-branes  provide the additional gauge bosons to form the adjoint of $SU(N_{D6})$. However our states are different. One reason is simply that for single centred Taub-NUT there is a harmonic two-form but no non-trivial two-cycle. More generally one sees that the soliton  profile is $A_+\sim \sum K_I\nu^I_+$ and  $0\le  K_I\le 1$ with equality {\it iff} $\vec x=\vec x_I$. Thus the $I$-th soliton is peaked at the $I$-th pole and furthermore vanishes at all the other poles. This means that the states we have found do not correspond to M2-branes which are wrapped on the non-trivial two-cycles. Rather our states are trapped at the poles, as discussed above. As such they are naturally associated to fundamental representations of the bulk enhanced gauge group, providing charged states of the bulk $SU(N_{D6})$ gauge group. From the point of view of the M2-brane worldvolume theory the wrapped M2-brane states arise as kink-like solitons, interpolating between pairs of poles as in \cite{Abraham:1992vb}.

\section*{Acknowledgements}
We would like to thank C. Bachas, C. Pagageorgakis, S. Schafer-Nameki, M. Schmidt-Sommerfeld and for discussions and H. Liu for collaboration on  \cite{LambertLiu}.  N. Lambert was supported in part by STFC grant
 ST/P000258/1 and M. Owen is supported by the STFC studentship ST/N504361/1.

\end{document}